\documentclass[%
 reprint,
 amsmath,amssymb,
 aps,
]{revtex4-2}

\usepackage{graphicx}% Include figure files
\usepackage{dcolumn}% Align table columns on decimal point
\usepackage{bm}% bold math
\usepackage{hyperref}% add hypertext capabilities
\usepackage[utf8]{inputenc} % allow Unicode characters like α
\usepackage{graphicx}% Include figure files
\usepackage{dcolumn}% Align table columns on decimal point
\usepackage{bm}% bold math
\usepackage{enumitem}
\usepackage{array}
\usepackage{siunitx}
\usepackage{adjustbox}
\usepackage{setspace}
\usepackage{dcolumn}
\usepackage{comment} 
\usepackage{hyperref}
\usepackage{gensymb}

\usepackage{threeparttablex}

\usepackage{booktabs} 
\usepackage{makecell} % add this in the preamble

\usepackage{placeins}

\usepackage{multirow}
\usepackage{tabularx}  % Automatically adjusts column widths
\usepackage{graphicx}  % For resizing tables
\graphicspath{{./figures/}}% Required for inserting images
\usepackage{url} %
\begin{document}

\preprint{APS/123-QED}
% \title{Data-Driven Uncertainty Quantification for Nuclear Reaction Parameters via Multi-Objective Optimization}
% \title{Data-Driven Uncertainty Quantification of Statistical Model Parameters via Multi-Objective Optimization}

\title{New approach for the quantification of uncertainties in reaction modeling via data-driven multi-objective optimization}
% \title{First ever set of uncertainty-quantified neutron resonance spacings for inaccessible experimentally unstable nuclei through data-driven multi-objective optimization of model parameters.}
\author{N. Dimitrakopoulos$^1$}
\email{dimit2n@cmich.edu}

\author{G. Perdikakis$^1$}
 \email{perdi1g@cmich.edu}
\author{F. Montes$^2$ }
 \author{P. Gastis$^3$}
 \author{S. A. Kuvin$^3$}
 \author{H. Y. Lee$^3$}
 \author{P. Tsintari$^2$}
 \author{A. V. Voinov$^4$}
\affiliation{%
$^1$Department of Physics, Central Michigan University, Mt. Pleasant, MI 48859, USA  }%

\affiliation{$^2$Facility for Rare Isotope Beams, Michigan State University, East Lansing, MI 48824, USA }%

\affiliation{
 $^3$Los Alamos National Laboratory, Los Alamos, New Mexico 87545, USA% with \\
}%

\affiliation{
 $^4$Department of Physics \& Astronomy, Ohio University, Athens, OH 45701, USA
}%

\date{\today}% It is always \today, today,

\begin{abstract}
We introduce a new multi-objective optimization approach to determine uncertainty-quantified nuclear reaction parameters in the Hauser-Feshbach framework. By simultaneously accounting for all available data across multiple reaction channels we capture parameter correlations and estimate data-driven uncertainties. We implement in the Ni-Ge region yielding uncertainty-quantified model parameters for both stable and unstable isotopes. We estimate resonance spacings for nuclei beyond experimental reach and validate our method by calculating measured cross-sections and nuclear level densities outside our optimization region. 

\end{abstract}
\maketitle
\section{Introduction}
Uncertainty quantification in reaction modeling has been recently identified as a key point of focus for the nuclear physics community \cite{Pruiit,Pruiit2,King,whitehead}, particularly in relevance to nuclear astrophysics data needs and the associated estimates of reaction yields involving isotopes off-stability where experimental data are scarce \cite{Hebborn_2023,Rausher,Bayessian_UQ}. For such applications, in which the Hauser-Feshbach statistical model serves as the main framework for obtaining reaction rates and cross sections, the uncertainty quantification should involve all model parameters used to describe nuclear structure properties such as nuclear level densities (NLD), optical model potentials (OMP), and gamma-ray strength functions ($\gamma$SF). 
Traditional approaches  of adjusting global parameters to fit experimental data typically involve ad hoc parameter tuning of individual channels, neglecting potential correlations with competing reactions. Furthermore, uncertainty quantification is often based on the spread among predictions from different model combinations  \cite{McKay,Denisenkov}, reflecting model divergence rather than alignment with experimental constraints. Significant efforts have been taken recently to use Bayesian statistics to quantify uncertainties in the parameters of the Koning-Delaroche global fit (\cite{Pruiit2}), and to propagate parameter uncertainties to reaction rates and nucleosynthesis yields \cite{Martinet2025} using Monte Carlo techniques and known neutron capture rates. 

In this article, we propose a new and different approach that employs a many-objective optimization strategy to identify the admissible region of model parameter space consistent with multiple experimental data, through the construction of a Pareto front of non-dominated solutions. The main assumption under which uncertainty is quantified in this approach is that a given set of model parameter values can vary only as much as allowed to adequately describe all available experimental data involving a particular target isotope. 
Using an optimization framework to balance competitive objectives across multiple reaction channels enables us to constrain the nuclear inputs and define the admissible region of parameter space. The systematic exploration of trade-offs between competing parameter sets leads to improved agreement with experimental data across multiple reaction channels, while avoiding overfitting to individual reactions. Then, uncertainty is quantified from the distribution of parameter values within the accepted Pareto solution ensemble, which is subsequently used to seed a constrained stochastic exploration for uncertainty estimation within the admissible trade-off region

In our work, we specifically focus on the following key question: ``\textit{How can we best make use of available data that are typically abundant along the valley of stability, to estimate changes in the average nuclear structure (spacings of excited levels, nuclear level density) for moderately nucleon-deficient unstable isotopes"}? We demonstrate that available experimental data can be successfully used to estimate NLD model parameters  and assign corresponding uncertainties that reflect the quality of reproduction of the known cross-section data. These parameters, in turn, allow for consistent, uncertainty-quantified estimates of key properties, such as average level spacings (e.g., the average s-wave neutron resonance spacing $D_0$).

Therefore, our approach directly addresses the challenge of quantifying such critical benchmark quantities for NLDs, such as the neutron resonance spacings \cite{Mughabhab2018} that suffer from a notable scarcity of experimental data, are not available for isotopes further than one neutron away from stability, and are often the result of a tedious analysis process.  These limitations are particularly impactful for applications in nuclear astrophysics and stockpile stewardship that involve reactions with unstable isotopes \cite{Carlson2017}. Moreover, experimental techniques such as the Oslo method \cite{OSLO,OSL01} rely on using the $D_0$ values as a normalization point at the neutron separation energy for the extracted NLD. With the extension of the method to unstable nuclei \cite{beta_oslo,inverso-Oslo}, direct $D_0$ measurements are unavailable, and normalizations often rely on systematics or data from neighboring nuclei \cite{Larsen,Liddick,Spyrou,Brits}. While mitigating the absence of strong experimental constraints, these approaches do not guarantee a realistic uncertainty quantification for the parameters. 

We show in this work that we can improve significantly over these approaches. In recent years, machine learning techniques have been employed to globally estimate NLD parameters and analyze systematics of resonance spacings with uncertainties \cite{ML_FGM,Bormans_2023}. Our approach advances upon previous works by proposing a method to extract ranges of acceptable level density values at the neutron separation energy $a(S_n)$. We couple a Hauser-Feshbach nuclear reaction code (TALYS 2.0 \cite{TALYS-2.0}) with two evolutionary optimizationn algorithms: the multi-Objective Non-dominated Sorting Genetic Algorithm II (NSGA-II \cite{NSGA2}) and the many-objective Reference Vector guided Evolutionary Algorithm (RVEA \cite{RVEA}), available from the pymoo \cite{pymoo} python library package. This framework constrains the model parameters by simultaneously fitting all available experimental data across the relevant reaction channels, and the resulting Pareto-optimal ensemble is subsequently used to initialize a constrained Metropolis \cite{Metropolis} sampling step, providing a denser exploration of the admissible trade-off region and enabling the construction of the uncertainty bands reported in this work.

% We couple the Hauser–Feshbach reaction code TALYS 2.0 \cite{TALYS-2.0} with two evolutionary optimization algorithms: the multi-objective Non-dominated Sorting Genetic Algorithm II (NSGA-II \cite{NSGA2}) and the many-objective Reference Vector guided Evolutionary Algorithm (RVEA \cite{RVEA}), both implemented in the pymoo \cite{pymoo} Python library. This framework constrains the model parameters by simultaneously fitting all available experimental data across the relevant reaction channels.

In this article, we report for the first time on a set of uncertainty-quantified level density model parameters for 10 unstable and 6 stable medium-mass isotopes, ranging from nickel to germanium, on both sides of the valley of stability. Furthermore, we supplement the NLD description with uncertainty-quantified OMP parameters for neutrons, protons, and alpha particles. Our uncertainty estimates are based on the application of our optimization procedure to all neutron-, proton-, and alpha-induced reactions on $^{68}$Zn for which cross-sections are available in the literature. The final product of our work that is presented here supplements the existing literature with average neutron resonance spacings, optical potential parameters, and level density model parameters estimated based on this analysis. For several of these nuclei, corresponding to isotopes a few neutrons away from stability, our work provides the first and only uncertainty-quantified estimate of level spacings available in the literature.

\section{Implementation of the Multi-Objective Optimization}
Multi-Objective Evolutionary Algorithms (MOEAs) \cite{MOEA} are a class of optimization techniques designed to solve problems with multiple competing objectives. These algorithms aim to identify a set of solutions that represent optimal trade-offs between the conflicting objectives. MOEAs evolve solutions over multiple generations to progressively approximate the Pareto-optimal front. A solution is considered Pareto-optimal if it is not dominated by any other solution in the feasible search space; in other words, there is no other solution that strictly improves all objectives.
A solution $x_1$ dominates $x_2$ if it is no worse in all objectives and strictly better in at least one:
\begin{equation}
x_1 \prec x_2 \iff 
\begin{aligned}
    &\forall i, f_i(x_1) \leq f_i(x_2) \\
    &\exists j, f_j(x_1) < f_j(x_2)
\end{aligned}
\end{equation}
where $f(x)$ represents the objective functions. These are in the most general case subject to a number (K) of constraints  $g_k(x)$ 
\begin{equation} g_k(x) \leq 0, \quad k = 1, \dots, K  \end{equation}

In this study, the many-objective optimization was carried out in two stages to manage the large search space. In the first stage, the NSGA-II algorithm was used as a multi-objective optimization to fit the elastic scattering angular distributions for protons and alpha particles at five incident energies each. This stage focused exclusively on the OMP parameters and identified a pool of OMP parameter sets that physically reproduce the elastic channel data. Using the physically meaningful parameter ranges determined for OMPs in the preceding step, in the second stage, the RVEA algorithm optimized both the OMP and NLD statistical model parameters, aiming to reproduce the measured cross sections for all reaction channels. These measured cross sections were used as the objectives in this second stage of the optimization process. The structure of the optimization codes is illustrated in Fig. \ref{fig:NSGA2_Schem} and  they are described in detail in the Supplemental Material \cite{suppM} (see also references \cite{Gabor,Bootstrap} therein).

The optimization framework was implemented in Python 3 using the pymoo \cite{pymoo} package. The TALYS 2.0 code \cite{TALYS-2.0} was used to perform statistical model calculations of neutron-, proton-, and alpha-induced reaction cross sections on $^{68}$Zn, when experimental data were available in the literature. The discrepancies between the calculated and experimental cross-sections were quantified using the reduced chi-square ($\chi_\nu^2$) values, which served as the minimization objectives in our optimization.

The optimization included a total of 29 observables, comprising 19 reaction channels and 10 elastic scattering angular distributions, corresponding to five proton-induced and five alpha-induced measurements at different incident energies. Specifically, the reactions considered are: for neutron-induced reactions, $(n,\text{Tot})$, $(n,n_0)$, $(n,n')$, $(n,n_3)$, $(n,p)$, $(n,\alpha)$, $(n,\gamma)$, as well as $(n,\gamma)_G$ and $(n,\gamma)_M$ activation data; for proton-induced reactions, $(p,\text{nonelastic})$, $(p,n)$, $(p,2n)$, $(p,3n)$, $(p,\alpha)$, $(p,2p)$, $(p,\gamma)$, and $(p,x)^{64}\text{Cu}$; and for alpha-induced reactions, $(\alpha,n)$ and $(\alpha,3n)$. For reactions such as $(p,n)$, $(p,2n)$, and $(p,2p)$, where many discrepant experimental data exist in the literature, the recommended data from the EXFOR library \cite{exfor} were used to compute the reduced chi-squared values.

\begin{figure}[tbh]
    \includegraphics[width=.99\columnwidth]{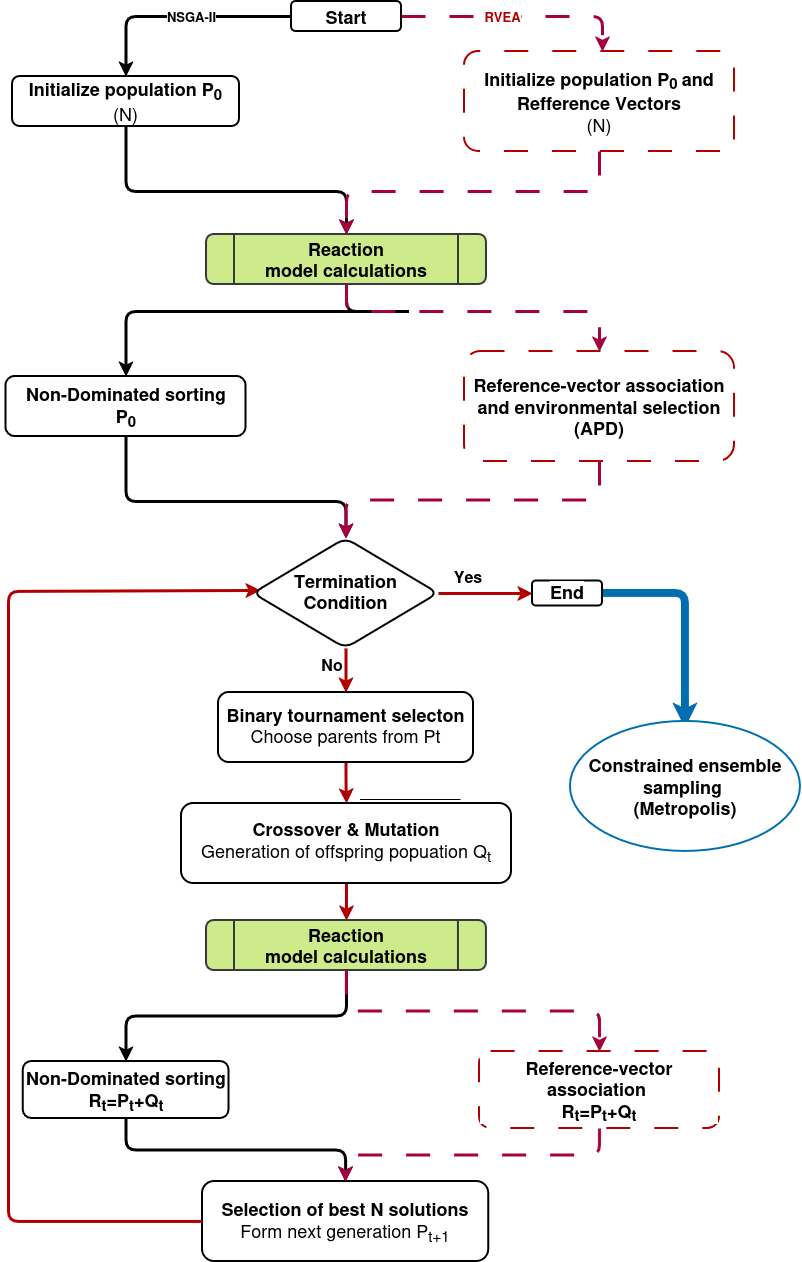} 
     \caption{
     Flow diagram illustrating the optimization workflow for the NSGA-II and RVEA algorithms. The parent population at generation $t$ is denoted by $P_t$, the offspring population by $Q_t$, and the combined set by $R_t$. Both algorithms follow the variation and evaluation steps, while differing in the environmental selection: NSGA-II uses dominance ranking and crowding distance, whereas RVEA utilizes the reference vector association and the angle-penalized distance (ADP). Dashed lines highlight how the flow diagram differs for the second-stage optimization using RVEA. The final Pareto-optimal ensemble is subsequently used to initialize the constrained Metropolis sampling procedure employed for uncertainty quantification.}
    \label{fig:NSGA2_Schem}
\end{figure}

\section{Theoretical modeling and parameter optimization}

Hauser-Feshbach calculations were performed using spherical OMPs, based on the Koning-Delaroche global parameterization \cite{KONING2003231} for protons and neutrons and the potentials of Avrigeanu et al. \cite{Avrigeanu} for $\alpha$-particles. Both parameterizations were adjusted during the optimization. The optical-model scattering problem was solved using the ECIS code as implemented in TALYS. For deformed nuclei, ECIS was used in coupled-channels mode with deformation parameters taken from the RIPL-3 library \cite{RIPL3}, while for spherical
nuclei the calculation reduces to the one-step Distorted-Wave Born approximation (DWBA).
% For all deformed nuclei, coupled-channel calculations were performed using deformation parameters available on RIPL-3 \cite{RIPL3} library to account for collective excitations.
The Gilbert-Cameron model \cite{CTM} was used for the calculation of all the NLD, and the $\gamma$SF function was modeled using the Skyrme HFB+QRPA approach \cite{GORIELY}.

% Forty-two parameters were optimized in this study, namely the radius $(r_v)$ and diffuseness $(a_v)$ parameters of the OMP for the protons, neutrons, and alphas, both the $\alpha$ level density parameter at the neutron separation energy $a(Sn)$, and the asymptotical value $\tilde{a}$ for the 16 nuclei involved in our calculations (see Fig. \ref{fig:chart}), the pre-equilibrium partial level densities for the isotopes  $^{67,68,69}$Ga, and finally the spin cut-off parameter ($\sigma^2$). 

\begin{figure}[htbp]
    \includegraphics[width=0.7\columnwidth]{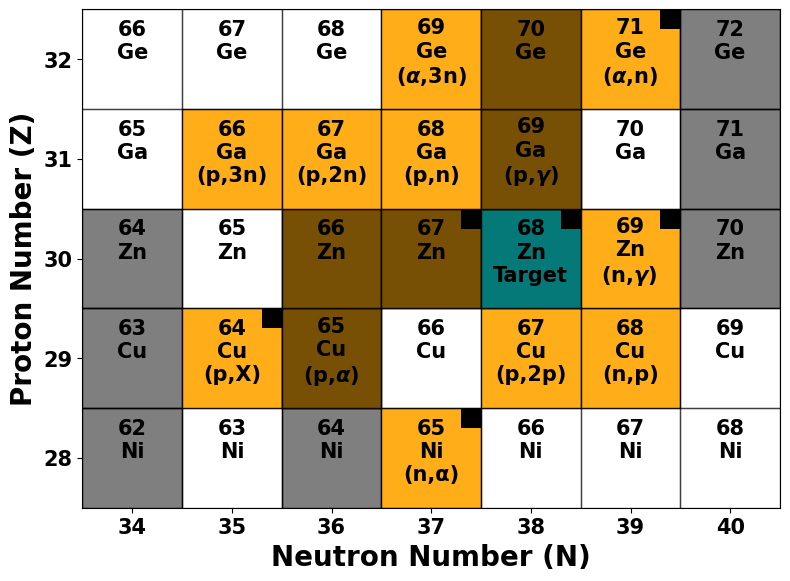} 
     \caption{ Segment of the chart of nuclei with the isotopes involved in our work. The target nucleus, $^{68}$Zn, is highlighted in teal, while orange boxes indicate nuclei whose level density parameters were optimized in our calculations. Dark shading indicates stable isotopes. The specific reactions used in the optimization are included for each residual nucleus. A black square in the top-right corner indicates availability of experimental s-wave resonance spacingnucleis ($D_0$).} 
    \label{fig:chart}
\end{figure}

A total of fifty-three parameters were optimized in this study. These include the OMP parameters for neutrons, protons, and alpha particles, the level density parameters, the pre-equilibrium partial level densities, and the spin cut-off parameter. For the OMPs, we optimized the real volume central parameters ($r_v$, $a_v$, $V_1$) for all three projectile types. For protons and alphas, the availability of elastic-scattering angular distributions allowed us to also constrain the imaginary volume central parameters ($r_w$, $a_w$, $W_1$), providing improved sensitivity to the absorptive part of the potential. For the proton OMP, the imaginary surface parameters ($r_{wd}$, $a_{wd}$, $D_1$) were included in the optimization, whereas for alpha particles the surface absorption was kept fixed because the high Coulomb barrier suppresses low-energy surface absorption. The remaining parameters include the level density parameters $a(S_n)$ and $\tilde{a}$ for the 16 nuclei involved in the calculations (see Fig.~\ref{fig:chart}), the pre-equilibrium partial level densities for $^{67,68,69}$Ga, and the spin cut-off parameter $\sigma^2$.

% To optimize $r_v$ and $a_v$ we adjusted the corresponding \textit{rvadjust} and \textit{avadjust} scaling parameters in TALYS. 

To optimize the 18 OMP parameters used in this work, we employed the corresponding adjust scaling keywords available in TALYS (e.g., \textit{rvadjust, avadjust, v1adjust}, etc.). 
% The next 32 optimization parameters were the $\alpha(Sn)$ and $\tilde{\alpha}$, for the sixteen nuclei involved in the calculation. These are described by the Ignatyuk formula :
The next 32 optimization parameters were the $\alpha(Sn)$ and $\tilde{\alpha}$  values for the sixteen nuclei involved in the calculation. These parameters enter the Gilbert–Cameron composite NLD model, in which a constant-temperature description is used at low excitation energies and is smoothly matched to a Fermi-gas formulation at higher energies. In this framework, the energy dependence of the Fermi-gas level-density parameter is given by the Ignatyuk prescription,
\begin{equation}
a(S_n)=\tilde{a} (1 + \delta W( \frac{1-e^{-\gamma (S_n-\Delta)}}{S_n-\Delta} ))
\label{eq:ignatyuk}
\end{equation}
where $\delta W$ is the shell correction energy defined as the difference between the experimental mass and the mass predicted by the liquid drop model \cite{MYERS19661}. $\gamma$ is the damping parameter, and $\Delta$ is an empirical energy shift; in this work, it is equal to the sum of the pairing energy and the pairing shift. Unlike traditional approaches \cite{Voinov2007} that adjust $\tilde{a}$ and $\Delta$  to calculate $\alpha(Sn)$ internally , we directly optimized both parameters and used  eq. \ref{eq:ignatyuk} to solve for $\gamma$, providing explicit control over lev
el density at $S_n$.
\begin{equation}
\gamma=-\frac{1}{S_n-\Delta}ln\left[ 1-\frac{S_n-\Delta}{\delta W}\left(\frac{\alpha(S_n)}{\tilde{\alpha}}-1 \right)\right]
\label{dampingpar}
\end{equation} 
To improve the description of the (p,Xn) channels leading to the formation of $^{67}$Ga, $^{68}$Ga, and  $^{69}$Ga at incident proton energies above 20 MeV, where pre-equilibrium and multi-step pre-equilibrium contributions become significant, we chose to optimize the partial pre-equilibrium level density parameter ($g$) of the two-component exciton model \cite{2compExModel} for the above nuclei. This was implemented by adjusting the \textit{gadjust} factor, to scale the global $g$ values of Koning and Duijvestijn \cite{KONING200415}.

The final optimization parameter was the width of the gaussian distribution for the angular momentum dependence of the NLD, through the so-called spin cut-off parameter $(\sigma^2)$, given in TALYS by the equation:
\begin{equation}
\sigma^{2}(E_x) = R_\text{spincut} \cdot 0.01389 \cdot \frac{A^{5/3}}{\tilde{a}} \cdot \sqrt{aU}
\label{ignatyuk}
\end{equation}
where $R_\text{spincut}$ is a global multiplier (the one we optimized), $A$ is the mass number of the nucleus, and $U$ is the excitation energy.

\subsection{Parameter exploration limits}

The OMP parameters were systematically varied over broad multipliers ranging from 0.2 to 1.8 in Stage 1 (NSGA-II). This broad search space was feasible due to the fast elastic-scattering calculations provided by TALYS/ECIS. The complete ranges for all the OMP scaling parameters are presented in detail in Table II of the Supplemental Material \cite{suppM}, along with the initial ranges of all parameters used to sample the first generation of solutions during both optimization processes.

The spin cut-off parameter multiplier was varied between 0.5 and 2.0, while the values of the pre-equilibrium partial level density multiplier ($g$) that we explored ranged between 0.5 and 1.6.

For nuclei with available experimental data on their s-wave resonance spacings, the $a(S_n)$ parameter was varied within $\pm 10\%$ from the values reported in RIPL-3 (\cite{RIPL3}). For all the other nuclei lacking experimental data we adopted a range of $\pm 20\%$ around the systematic value. This systematic value is calculated by substituting the $\tilde{a}$ value derived from systematics into ignatyuk formula (eq. \ref{eq:ignatyuk}), providing explicit control over level density at $S_n$.

 The formula for $\tilde{a}$, based on systematics, is expressed as:
\begin{equation}
    \tilde{a}=d_1 A+d_2 A^{\frac{2}{3}}
\end{equation}
where $d_1, d_2$ are obtained from the global parametrization of Koning and Delaroche.

Finally, the ranges for the asymptotic parameter ($\tilde{\alpha}$) can be obtained by demanding the damping parameter to remain positive (effectively demanding the natural logarithm to be negative). Consequently,  the upper and lower limits for the  $ \tilde{a}$ value are :
\begin{equation}
    \frac{a_{min}(Sn)}{\frac{Sn-\Delta}{\delta W}+1} < \tilde{a}<a_{max}(Sn)
    \label{eq:atilde}
\end{equation}

\subsection{Constraints}
To restrict the parameter search space, we applied sets of inequality and equality constraints to penalize non-physical solutions. For example, the available experimental data for the s-wave resonance spacing ($D_0$)  were utilized as constraints, requiring that our calculated $D_0$ values lie within the experimental uncertainty. In cases where the difference exceeded the experimental uncertainty, an exponential penalty function was applied, ensuring that larger deviations resulted in stronger penalties.

We imposed additional constraints on the parameter space to prevent recalculations of the shell correction value for all nuclei involved,  which would otherwise be required when physically inconsistent parameter combinations lead to negative damping parameters.
These constraints stem from Eq. \ref{dampingpar}, as we are working within a defined range for $a(Sn)$ and $\tilde{a}$. If $\tilde{a}$ is chosen near its upper limit, while $a(\text{Sn})$ is selected at the lower edge of the allowed range, there is a risk that the damping parameter becomes negative. 
To exclude such combinations of parameters from the search space, a penalty factor of 1e6 was applied whenever the optimization process resulted in negative damping parameters.

The remaining 18 constraints required that the $\chi^2_\nu$ values between experimental and calculated data be lower than those obtained using the global TALYS parametrization. The $\chi^2_\nu$ values were evaluated using an effective cross-section uncertainty, in which the reported y-axis uncertainties were combined with the contribution from x-axis uncertainties through standard error propagation. For each dataset, the $\chi^2_\nu$ obtained with the default parametrization was used as a reference, and candidate solutions were penalized only if their $\chi^2_\nu$ exceeded this reference value by more than 2.5. This tolerance was introduced to accommodate undocumented systematic effects and to ensure a fairer comparison across datasets, since the $\chi^2_\nu$ evaluation includes only statistical uncertainties and systematic uncertainties were rarely reported in the experimental literature. When energy uncertainties were not provided, a minimum value of 5 keV was assigned solely for the calculation of the effective uncertainty entering $\chi^2_\nu$. These choices provide a more uniform treatment of heterogeneous datasets while avoiding the over-penalization of physically reasonable solutions arising from incomplete systematic uncertainty information. If a candidate solution exceeded this acceptance threshold, a penalty was applied proportionally to the fractional contribution of the corresponding reaction channel to the total cross section. For example, deviations in reaction channels with cross-sections around 1 barn were penalized by multiplying their difference by 1000, while for cross-sections below 1 mb, the penalty was simply the reduced chi-square value without any additional multiplier.

\subsection{Ensemble sampling for uncertainty quantification}

The evolutionary optimization produces an ensemble of Pareto optimal parameter sets that simultaneously satisfy all experimental constraints across the reaction channels. This ensemble determines in a course manner the admissible region of parameter space consistent with the available data and the imposed objective thresholds. To obtain a finer characterization of the variability permitted within this region, we subsequently applied a constrained Metropolis \cite{Metropolis} sampling procedure initialized from each Pareto optimal solution. Starting from every member of the Pareto set, an independent Markov chain was constructed by proposing trial parameter vectors through Gaussian perturbations around the current state, with proposal widths determined from the parameter spreads of the RVEA ensemble. Since the Metropolis algorithm requires a scalar acceptance criterion, a global energy function was defined as the sum of the reduced chi square objectives across all reaction channels. Candidate parameter sets were accepted only if they obeyed the constraints of eq. \ref{eq:atilde} and also remained within the objective bounds established by the Pareto solutions: each individual objective was required to stay below the maximum value attained among the Pareto optimal ensemble, and the total summed objective was required to remain below the maximum summed value observed across the Pareto set. Within this admissible region, trial steps yielding improved chi-square were always accepted, while occasional uphill moves were permitted with the standard Metropolis probability to ensure local exploration around each Pareto optimal solution. This constrained sampling step was then used to generate the uncertainty bands and parameter distributions reported throughout this work.

\section{Optimization results and uncertainty quantification}

\begin{figure*}[htpb]
    \includegraphics[width=.99\textwidth]{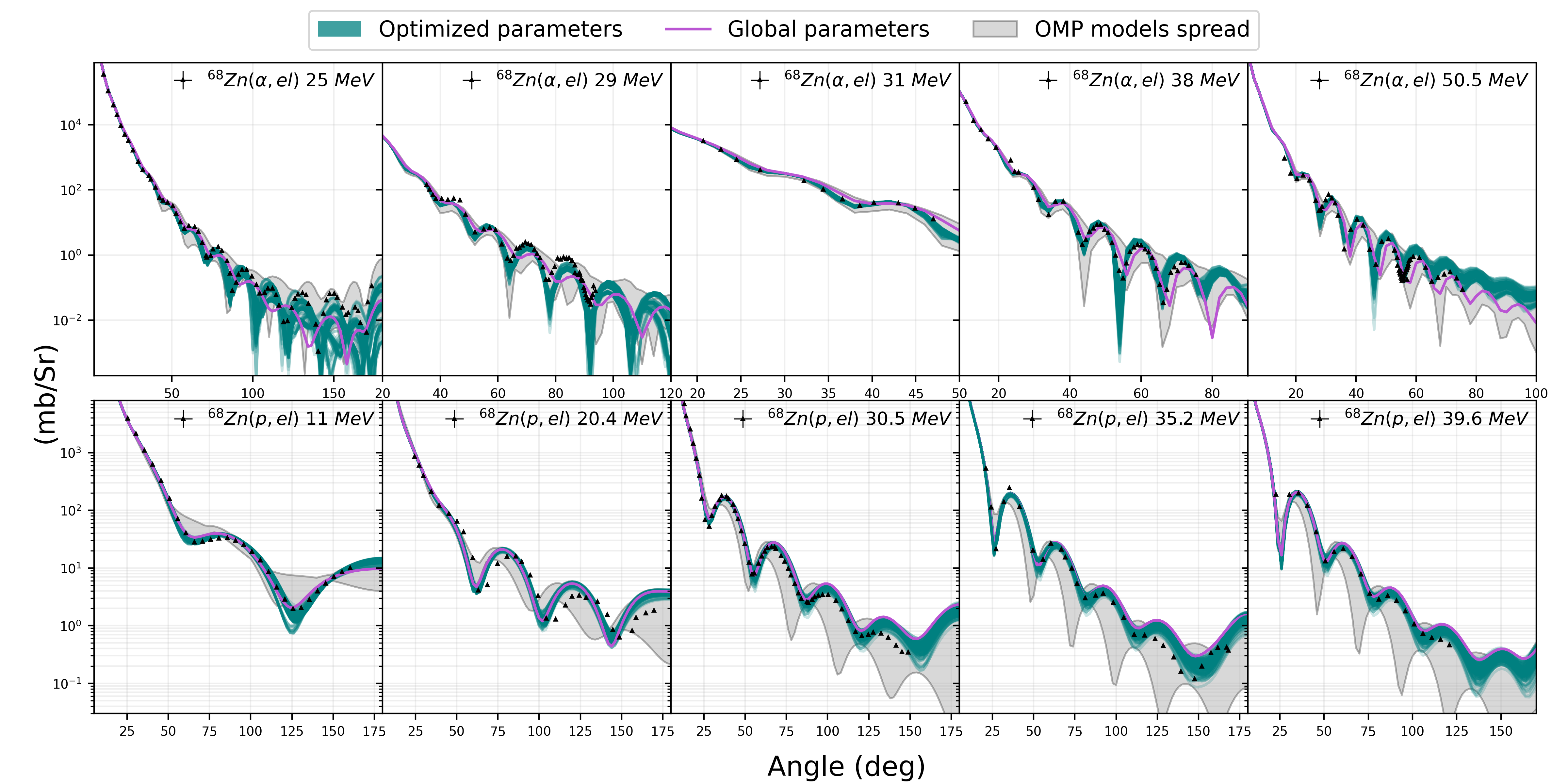} 
    \caption{Differential elastic cross sections for $^{68}Zn$ using alpha particles (top row) with incident energies ranging from 25 to 50.5 MeV, and protons (bottom row) with energies between 11 and 39.6 MeV. 
    The teal band represents the Pareto-optimal solutions obtained from this work.
    The teal band represents the final uncertainty range obtained from Metropolis sampling initialized from the Pareto optimal solutions of this work.
    Purple curves show results from default TALYS calculations (ECIS). The gray band represents the spread of predictions using all available alpha particle optical model potentials and, for protons, both spherical and microscopic JLM models.}
    \label{elastic}
\end{figure*}

In Fig. \ref{elastic} we present the differential elastic scattering cross sections for alpha particles on $^{68}$Zn (top row, 25–50.5 MeV) and for protons on $^{68}$Zn (bottom row, 11–39.6 MeV). The teal bands correspond to the final accepted OMP parameter ensemble obtained in this work, while the purple curves represent calculations using the global parametrization of Koning and Delaroche. The gray bands indicate the spread of predictions from all available OMP models for alphas \cite{Avrigeanu,Avrigeanu1994,MCFadden,Nolte,DEMETRIOU2002}, and both spherical and microscopic JLM models \cite{JML1,JLM2} for protons. Our optimized OMPs reproduce the experimental data across all energies better than the global OMPs. In the forward-angle region, where the real part of the potential dominates, the agreement is excellent, capturing both the magnitude and shape of the data. At larger scattering angles, the absorptive term becomes more significant, leading to increased uncertainty in our predictions, but the overall agreement remains superior to the global parametrization.

\begin{figure*}[htbp]
    \includegraphics[width=.99\textwidth]{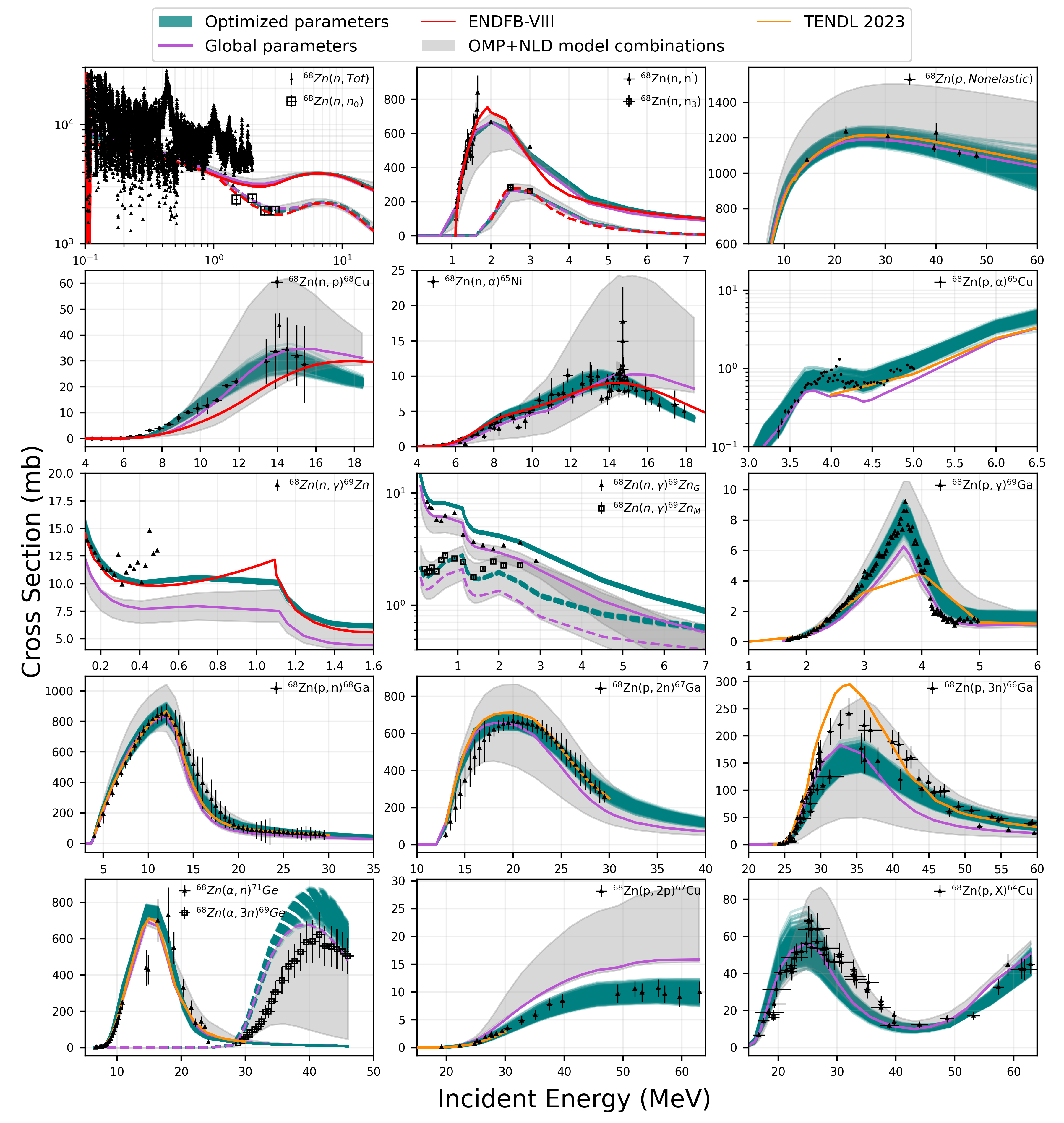} 
     \caption{Calculated cross-sections for all reactions on $^{68}$Zn where experimental data are available. 
     % The Pareto-optimal solutions from this work are represented in teal color.
     The results from this work are represented in teal color, showing the uncertainty band obtained from the ensemble of accepted parameter sets after the many-objective optimization and the subsequent constrained Metropolis sampling.
     The results of a ``default" TALYS calculation are shown with a purple line. The ENDF/B-VIII.0 and TENDL-2023 evaluations are indicated with red and orange lines respectively. The width of the teal color band shows the range of results obtained with the optimization approach. Dashed areas of the same color denote a second reaction channel in the same plot. Black symbols are used for the experimental data and the corresponding uncertainties. The results obtained by varying level density and gamma-ray strength function models is depicted with the gray shaded area. Overall, our calculations improve on the agreement compared to the evaluated libraries and unoptimized calculations with TALYS. The optimized results yield a substantially more constrained uncertainty range compared to the spread from all combinations of OMP and NLD models (gray band)}.       
    \label{fig:Cross_sections}
\end{figure*}

The calculated cross sections from our work are presented in Fig. \ref{fig:Cross_sections}, together with the corresponding uncertainty envelope from the optimized ensemble. Our results demonstrate a nearly universally better agreement with the data when compared with calculations using the global parameterization suggested by default in TALYS (purple). Similarly, our optimized parameters seem to perform generally better than the ENDF/B-VIII.0 \cite{ENDF8} (red) evaluation for neutron-induced reactions and the TENDL-2023 \cite{TENDL}  evaluation (orange) for proton- and alpha-induced reactions. Most importantly, our results provide a data-based quantified uncertainty band that we can correlate with specific model parameter uncertainties, and which is absent in single calculations or evaluated results. Compared to a typically used error band derived from all possible combinations of OMP and NLD models (for examples of use see e.g. \cite{beta_oslo}, \cite{Nikas}) (gray band), our optimization approach yields uncertainty bands more than five times narrower for some of the reactions we studied. These results highlight that uncertainty bands derived solely from varying different models may overestimate the true physical uncertainty, because they reflect the spread of model assumptions rather than the actual reaction uncertainty. In contrast, our data-driven optimization produces narrower bands that are directly correlated with parameter sensitivities, yielding a more realistic estimate of the reaction uncertainty.

The optimized model parameters and their associated data-based uncertainties are presented in Table \ref{tab:RESULTS_OMP} (OMP parameters) and Table \ref{tab:RESULTS} (NLD model parameters). In Table \ref{tab:RESULTS}, beyond the three parameters explicitly optimized for each isotope ($\alpha(S_n)$, $\tilde{\alpha}$, and $R_{\text{spincut}}$), we also report the remaining Gilbert–Cameron model parameters. These additional parameters were not independently optimized but are deterministically derived from the optimized quantities. In particular, the constant-temperature model parameters are obtained by enforcing continuity of the level density at the matching energy, while the damping parameter $\gamma$ is calculated from the optimized $\alpha(S_n)$ values using eq. \ref{eq:ignatyuk}. Kernel density estimation (KDE) was used to deduce the most probable values from each parameter distribution, and uncertainties were quantified using central 68\% credible intervals, as described in the Supplemental Material \cite{suppM}. There, along with a summary table of key statistical measures and additional discussion, we present the resulting distributions for each parameter optimized in this work.

In Table \ref{tab:RESULTS_OMP}, the real parts of the optical potentials ($r_v$, $a_v$, $V_1$) for all projectiles remain within 8\% of their default values. The imaginary components, however, exhibit larger deviations for protons and alpha particles, changes that are fully justified by the excellent agreement achieved with both the differential elastic-scattering data (Fig. \ref{elastic}) and the reaction cross sections (Fig. \ref{fig:Cross_sections}). For neutrons, the real part changed by less than 2\%, with the imaginary component fixed due to the absence of angular distribution data. This minor adjustment highlights that the optimization naturally preserved physically reasonable parameters.
The optimized statistical-model input parameters are summarized in Table \ref{tab:RESULTS}. For several nuclei, the resulting NLD curves remain close to the global parametrization within the extracted uncertainty envelope, indicating that the available experimental constraints do not require substantial modifications to the default description. For other nuclei, the optimization leads to input parameters that deviate more stronlgy from the global values; these cases are underlined in Table \ref{tab:RESULTS}. The resulting NLD curves and their associated uncertainty envelopes are shown for all nuclei in the Supplemental Material \cite{suppM}, together with a comparison to the global prescription. In addition, sizable deviations from the global parametrization were observed for the partial pre-equilibrium level-density multipliers $g$, which were adjusted to reproduce proton-induced data above 20 MeV, where multi-step processes become increasingly important. The optimized \textit{gadjust} factors were found to be $0.67 \pm 0.15$ for $^{67}$Ga, $1.43 \pm 0.01$ for $^{68}$Ga, and $0.97 \pm 0.02$ for $^{69}$Ga. A detailed discussion of these pre-equilibrium effects is beyond the scope of the present work.

% \begingroup
% \scriptspace=0pt
% \medmuskip=1mu
% \thickmuskip=2mu
% \begin{table}[htbp]
% \centering
% \renewcommand{\arraystretch}{1.30}
% \caption{A summary of all the OMP parameters optimized in this work.}
% \begin{tabular}{|c|c|c|c|}
% \cline{1-4}
% \textbf{Particle} &\boldsymbol{$\alpha$} &\textbf{p}&\textbf{n} \\
% \cline{2-4}
% \textbf{\textit{rv}}& 1.003    $^{+0.004}_{-0.000}$&0.987    $^{+0.004}_{-0.003}$&1.017    $^{+0.001}_{-0.001}$\\
% \textbf{\textit{av}}& 1.076    $^{+0.002}_{-0.002}$&1.052    $^{+0.023}_{-0.009}$& 0.998    $^{+0.001}_{-0.001}$\\
% \textbf{\textit{V1}}& 0.962    $^{+0.004}_{-0.000}$&1.006    $^{+0.001}_{-0.003}$& 1.000    $^{+0.002}_{-0.002}$\\
% \textbf{\textit{rw}}& 1.034    $^{+0.022}_{-0.004}$&1.022    $^{+0.002}_{-0.068}$&|\\
% \textbf{\textit{aw}}& \footnote{$-16.6149 \cdot$ ${V1_{\alpha}}^2$ + $31.14981 \cdot$ ${V1_{\alpha}}$ - $13.4742$} $^{+0.018}_{-0.018}$&0.557    $^{+0.030}_{-0.068}$&|\\
% \textbf{\textit{W1}}&  \footnote{$-64.1224 \cdot {av_p}^2 + 140.0454 \cdot {av_p} - 75.3494$}   $^{+0.068}_{-0.068}$&0.893    $^{+0.025}_{-0.059}$&|\\
% \textbf{\textit{rwd}}&|&1.003    $^{+0.044}_{-0.170}$&|\\
% \textbf{\textit{awd}}&|&0.987    $^{+0.003}_{-0.001}$&|\\
% \textbf{\textit{D1}}&|&1.034    $^{+0.086}_{-0.018}$&|\\
% \hline
% \cline{2-3}
% \end{tabular}
% \label{tab:RESULTS_OMP}
% \end{table}
% \endgroup

\begingroup
\scriptspace=0pt
\medmuskip=1mu
\thickmuskip=2mu
\begin{table}[htbp]
\centering
\renewcommand{\arraystretch}{1.30}
\caption{A summary of all the OMP parameters optimized in this work.}
\begin{tabular}{|c|c|c|c|}
\cline{1-4}
\textbf{Particle} &\boldsymbol{$\alpha$} &\textbf{p}&\textbf{n} \\
\cline{2-4}
\textbf{\textit{rv}}& 1.004    $^{+0.002}_{-0.002}$&0.986    $^{+0.008}_{-0.005}$&1.018    $^{+0.001}_{-0.002}$\\
\textbf{\textit{av}}& 1.077    $^{+0.002}_{-0.005}$&1.027   $^{+0.039}_{-0.004}$& 0.998    $^{+0.001}_{-0.001}$\\
\textbf{\textit{V1}}& 0.949    $^{+0.021}_{-0.091}$&1.005    $^{+0.003}_{-0.002}$& 1.001    $^{+0.003}_{-0.003}$\\
\textbf{\textit{rw}}& 1.043    $^{+0.028}_{-0.024}$&1.014    $^{+0.071}_{-0.047}$&|\\
\textbf{\textit{aw}}& \footnote{$-16.6149 \cdot$ ${V1_{\alpha}}^2$ + $31.14981 \cdot$ ${V1_{\alpha}}$ - $13.4742$} $^{+0.018}_{-0.018}$&0.574    $^{+0.089}_{-0.077}$&|\\
\textbf{\textit{W1}}&  \footnote{$-64.1224 \cdot {av_p}^2 + 140.0454 \cdot {av_p} - 75.3494$}   $^{+0.068}_{-0.068}$&0.907    $^{+0.072}_{-0.085}$&|\\
\textbf{\textit{rwd}}&|&1.023    $^{+0.158}_{-0.294}$&|\\
\textbf{\textit{awd}}&|&0.988    $^{+0.004}_{-0.005}$&|\\
\textbf{\textit{D1}}&|&1.048    $^{+0.102}_{-0.050}$&|\\
\hline
\cline{2-3}
\end{tabular}
\label{tab:RESULTS_OMP}
\end{table}
\endgroup

\begingroup
\scriptspace=0pt
\medmuskip=1mu
\thickmuskip=2mu
\begin{table*}[htbp]
\centering
\renewcommand{\arraystretch}{1.30}
\caption{A summary of all the statistical model parameters optimized in this work. Underlined isotopes correspond to nuclei for which the optimized inputs deviate most strongly from the  global parametrization}
\begin{tabular}{|c|c|c|c|c|c|c|c|c|}
\cline{1-9}
\textbf{Isotope} & \boldsymbol{$T $ $[MeV]$}& \boldsymbol{$E_0$ $[MeV]$}& \boldsymbol{$E_M$ $[MeV]$}& \boldsymbol{$\alpha(Sn)$ $[MeV^{-1}]$} & \boldsymbol{$\tilde{\alpha}$ $[MeV^{-1}]$} &\boldsymbol{$\delta W$ $[MeV]$ }$^*$ &\boldsymbol{$\gamma$ } & \boldsymbol{$\sigma(S_n)$ $[MeV]$} \\
\cline{2-9}
\underline{$^{65}$Ni} & 0.85 $^{+0.01}_{-0.06}$& 0.25 $^{+0.18}_{-0.04}$& 4.15 $^{+0.35}_{-0.68}$& 10.18 $^{+0.01}_{-0.01}$ & 8.10   $^{+0.12}_{-0.02}$ & 1.2290  & 0.576   $^{+0.250}_{-0.091}$& 3.60 $^{+0.01}_{-0.03}$\\
\underline{$^{64}$Cu} & 1.12 $^{+0.00}_{-0.00}$& 2.57 $^{+0.00}_{-0.00}$& 7.06 $^{+0.12}_{-0.01}$& 8.66  $^{+0.00}_{-0.00}$  & 8.72  $^{+0.10}_{-0.02}$  &-0.2575 &0.033   $^{+0.076}_{-0.011}$& 3.77 $^{+0.01}_{-0.02}$\\
\underline{$^{65}$Cu} & 1.07 $^{+0.01}_{-0.01}$& -0.77 $^{+0.04}_{-0.03}$& 7.91 $^{+0.09}_{-0.15}$& 8.78  $^{+0.05}_{-0.04}$  & 8.59  $^{+0.03}_{-0.03}$& 0.5176 & 0.055   $^{+0.025}_{-0.019}$& 3.92 $^{+0.01}_{-0.01}$\\
$^{67}$Cu & 0.93 $^{+0.03}_{-0.12}$& -0.37 $^{+0.44}_{-0.52}$& 6.59 $^{+0.95}_{-1.67}$& 10.34 $^{+0.19}_{-0.27}$  & 9.18  $^{+0.55}_{-0.32}$&1.4776 & 0.076   $^{+0.190}_{-0.026}$& 3.97 $^{+0.06}_{-0.13}$\\
$^{68}$Cu & 0.94 $^{+0.01}_{-0.02}$& -2.38 $^{+0.12}_{-0.13}$& 6.06 $^{+0.19}_{-0.35}$& 10.67 $^{+0.06}_{-0.010}$  & 9.89  $^{+0.16}_{-0.14}$& 1.7101& 0.050   $^{+0.018}_{-0.011}$& 3.70 $^{+0.04}_{-0.02}$\\
$^{66}$Zn & 1.06 $^{+0.03}_{-0.04}$& 0.02 $^{+0.17}_{-0.11}$& 10.29 $^{+0.21}_{-0.83}$& 9.60  $^{+0.34}_{-0.18}$  & 9.34  $^{+0.17}_{-0.26}$& 0.7972 &  0.034   $^{+0.114}_{-0.011}$&  3.86   $^{+0.05}_{-0.03}$ \\
$^{67}$Zn & 0.98 $^{+0.01}_{-0.00}$& -1.37 $^{+0.01}_{-0.03}$& 7.90 $^{+0.13}_{-0.15}$& 10.72 $^{+0.06}_{-0.05}$  & 9.46  $^{+0.14}_{-0.11}$  & 1.8464 & 0.090   $^{+0.017}_{-0.013}$& 3.63   $^{+0.02}_{-0.02}$ \\
\underline{$^{68}$Zn} & 1.03 $^{+0.00}_{-0.01}$& 0.02 $^{+0.02}_{-0.02}$& 9.35 $^{+0.18}_{-0.15}$& 10.08 $^{+0.03}_{-0.03}$  & 8.79  $^{+0.12}_{-0.07}$ & 1.9539& 0.108 $^{+0.014}_{-0.017}$& 4.02 $^{+0.02}_{-0.03}$\\
\underline{$^{69}$Zn} & 0.89 $^{+0.00}_{-0.01}$& -1.06 $^{+0.02}_{-0.01}$& 6.17 $^{+0.17}_{-0.16}$& 11.74 $^{+0.05}_{-0.03}$  & 8.86  $^{+0.15}_{-0.9}$ & 2.6427 & 0.192 $^{+0.017}_{-0.021}$& 3.84 $^{+0.02}_{-0.03}$\\
$^{66}$Ga & 1.12 $^{+0.01}_{-0.01}$& -3.37 $^{+0.03}_{-0.06}$& 8.06 $^{+0.12}_{-0.39}$& 9.40  $^{+0.05}_{-0.08}$  & 9.18  $^{+0.12}_{-0.11}$  & 0.3563& 0.0.53 $^{+0.158}_{-0.017}$& 4.99 $^{+0.02}_{-0.03}$\\
\underline{$^{67}$Ga} & 1.04 $^{+0.01}_{-0.03}$& -1.00 $^{+0.10}_{-0.04}$& 8.09 $^{+0.10}_{-0.36}$& 9.43  $^{+0.20}_{-0.08}$  & 9.20  $^{+0.04}_{-0.03}$  &1.0864& 0.024 $^{+0.033}_{-0.009}$ & 4.11 $^{+0.02}_{-0.02}$\\
$^{68}$Ga & 0.98 $^{+0.03}_{-0.10}$& -2.28 $^{+0.41}_{-0.15}$& 6.16 $^{+0.30}_{1.40}$& 9.93  $^{+0.59}_{-0.23}$  & 9.37  $^{+0.14}_{-0.10}$  &1.5641 &  0.036 $^{+0.97}_{-0.013}$& 4.00 $^{+0.07}_{-0.02}$\\
\underline{$^{69}$Ga} & 0.89 $^{+0.02}_{-0.05}$& -0.45 $^{+0.16}_{-0.06}$& 6.74 $^{+0.18}_{-1.17}$& 10.21 $^{+0.22}_{-0.15}$  & 8.81  $^{+0.92}_{-0.26}$  & 2.1811& 0.048 $^{+0.137}_{-0.016}$& 4.24 $^{+0.10}_{-0.17}$\\
$^{69}$Ge & 0.95 $^{+0.04}_{-0.04}$& -1.76 $^{+0.15}_{-0.17}$& 8.57 $^{+0.35}_{-0.5}$& 11.51 $^{+0.41}_{-0.38}$  & 10.95  $^{+0.39}_{-0.85}$  &2.2259&  0.019 $^{+0.61}_{-0.007}$& 3.69 $^{+0.14}_{-0.05}$\\
\underline{$^{70}$Ge} & 0.77 $^{+0.02}_{-0.08}$& 1.00 $^{+0.24}_{-0.07}$& 6.58 $^{+0.32}_{-1.09}$& 11.86 $^{+0.17}_{-0.18}$  & 9.62  $^{+0.21}_{-0.17}$  &2.2224& 0.230 $^{+0.193}_{-0.033}$& 4.28 $^{+0.02}_{-0.05}$ \\
\underline{$^{71}$Ge} & 0.90 $^{+0.00}_{-0.00}$& -1.69 $^{+0.01}_{-0.01}$& 8.38 $^{+0.04}_{-0.02}$& 12.29 $^{+0.03}_{-0.05}$  & 11.72 $^{+0.9}_{-0.09}$  &3.2202&0.016 $^{+0.002}_{-0.004}$& 3.60 $^{+0.01}_{-0.02}$\\
\hline
\end{tabular}
\begin{tablenotes}
\item[a] {\small $^*$ Shell correction values were used as inputs and remained constant through the optimization}
\end{tablenotes}
\label{tab:RESULTS}
\end{table*}
\endgroup

\section{Optimization Benchmarks}

\begin{figure}[htbp]
    \includegraphics[width=0.9\columnwidth]{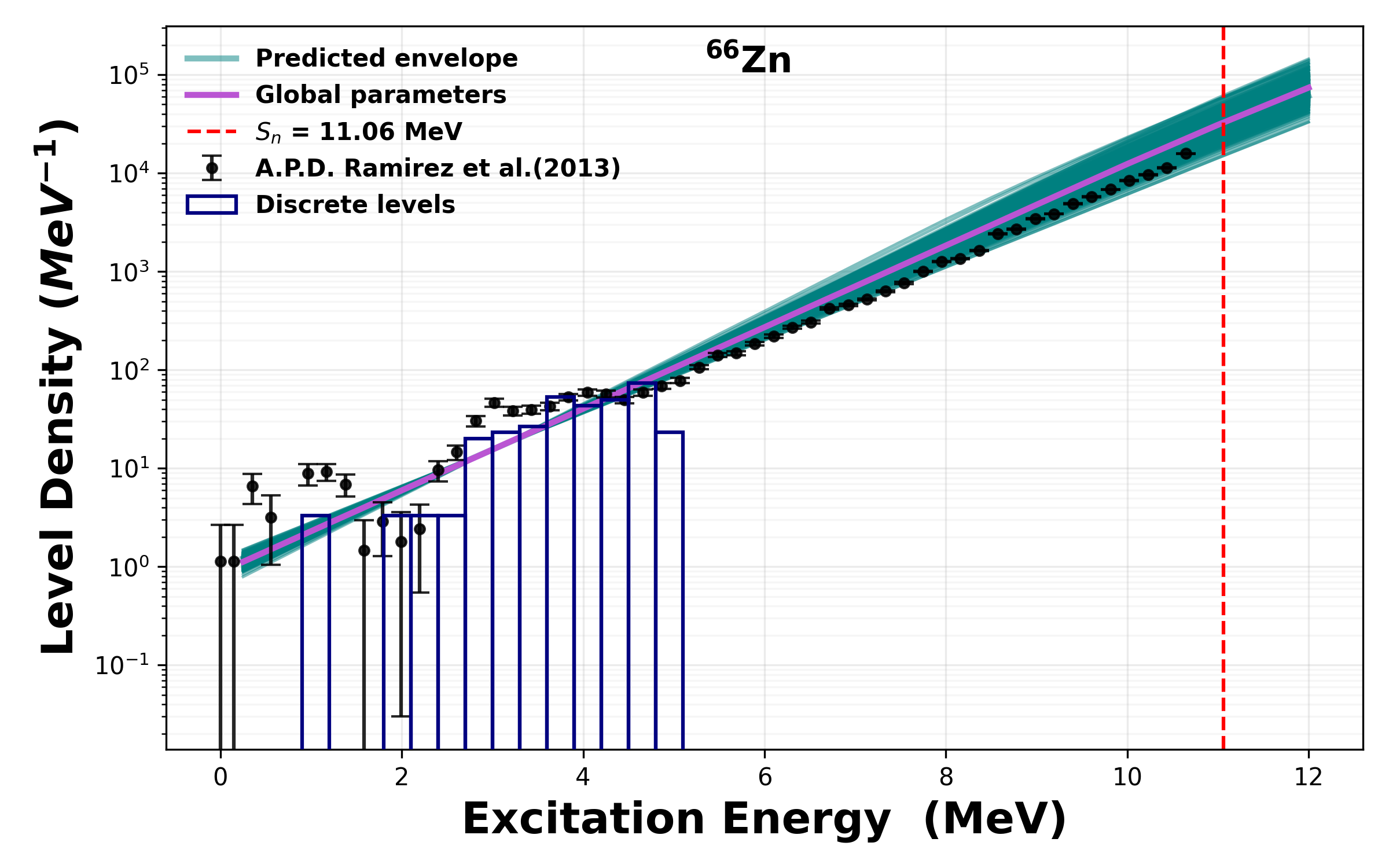}
    \includegraphics[width=0.9\columnwidth]{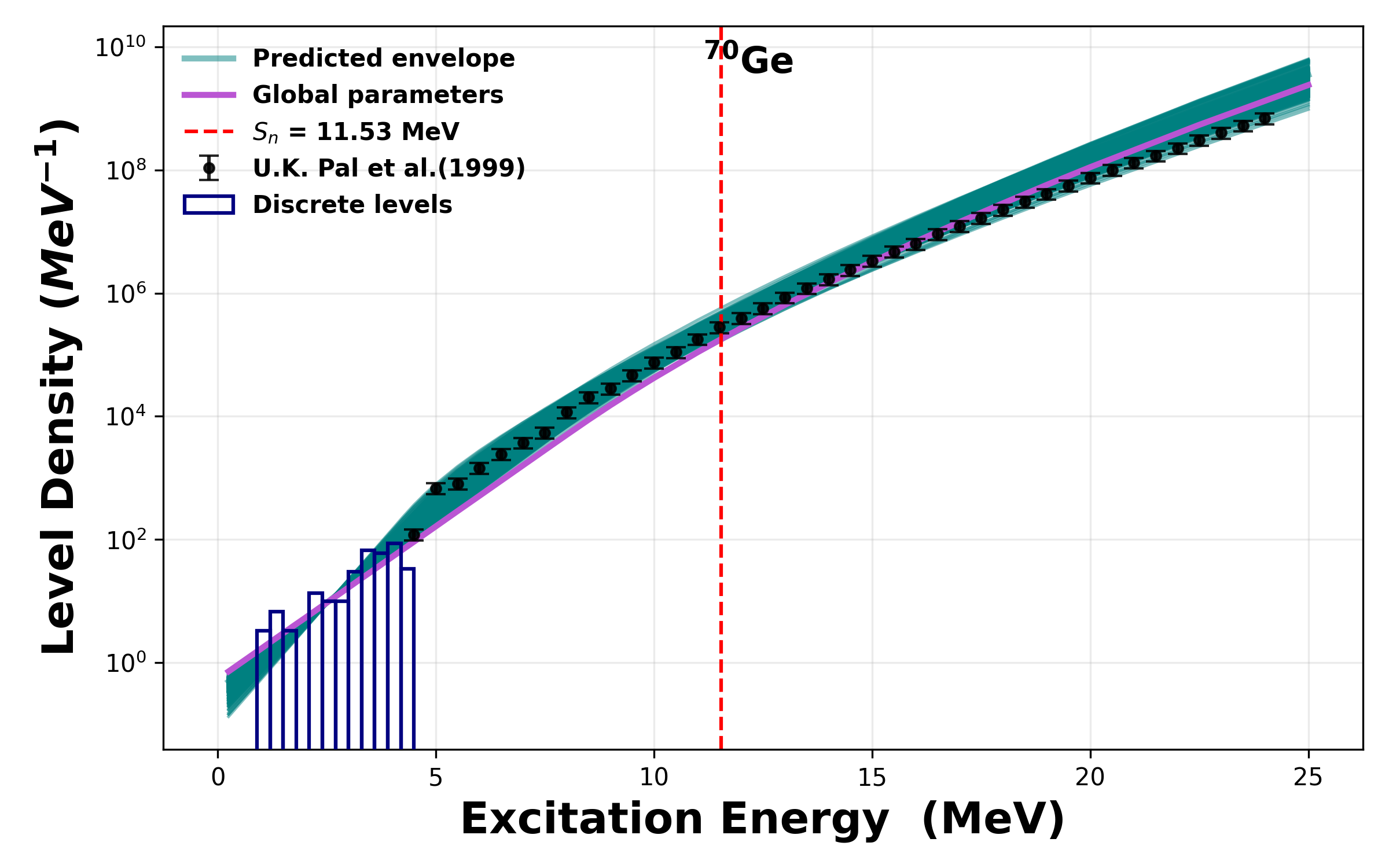}
     \caption{ Benchmark comparison of experimentally extracted nuclear level densities for $^{66}$Zn (top) and $^{70}$Ge 
(bottom). Experimental NLD data are shown as black symbols, while the low-energy discrete levels are indicated by dark-blue bars. The teal band represents the uncertainty envelope obtained from this work, while the calculations using the  global parametrization are shown in the purple curves. The neutron separation energy $S_n$is marked by the vertical dotted red line in each panel.} 
    \label{Predictions NLD}
\end{figure}

As an external validation of our uncertainty-quantification approach, we consider two nuclei involved in our calculations for which independent experimental NLD data are available in the CANDL database \cite{CANDL}. Importantly, these level-density data were not included among the objectives or constraints of the optimization. The benchmark nuclei are $^{66}$Zn and $^{70}$Ge; in both cases the NLDs were extracted using the evaporation technique and have been reported in Refs.~\cite{Voinov_66Zn,UKPal_1999}.
Fig. \ref{Predictions NLD} compares the experimental level densities for $^{66}$Zn (top panel) and $^{70}$Ge (bottom panel) with the uncertainty bands obtained from the optimized and sampled parameter ensemble. For reference, the predictions obtained with the default TALYS global parametrization are also shown. In both cases, the experimental level densities are reproduced within the extracted uncertainty envelope. For  $^{70}$Ge in particular, the optimized results provide a substantially improved description of the experimental NLD shape compared to the default parametrization, demonstrating that the present framework yields predictive constraints beyond the datasets directly included in the optimization.

\begin{figure*}[htpb]
    \includegraphics[width=.98\textwidth]{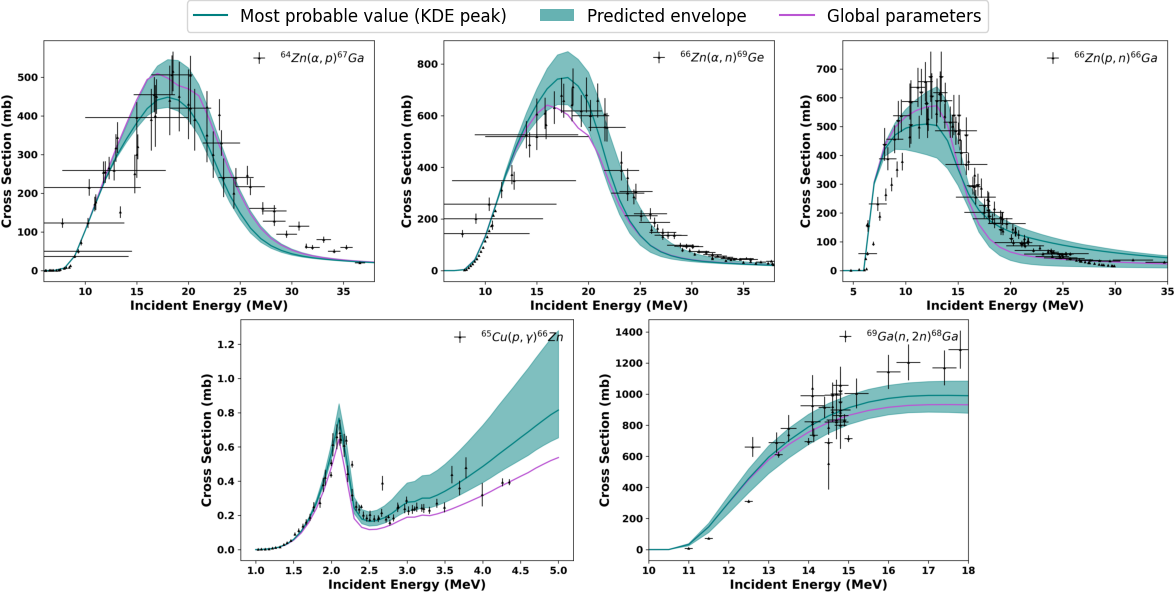}
    \caption{Predicted cross sections for five reactions that were not included in the optimization. The reactions shown (from top left to bottom right) are: $^{64}$Zn$(\alpha,p)^{67}$Ga, $^{66}$Zn$(\alpha,n)^{69}$Ge, $^{65}$Cu$(p,\gamma)^{66}$Zn, and $^{69}$Ga$(n,2n)^{68}$Ga. The opaque teal band represents the predicted envelope arising from the range of optimized NLD parameters relevant to each reaction. The teal line indicates the most probable prediction, corresponding to the KDE s of the optimized parameter distributions, and is compared to the TALYS default calculation (purple) for all reactions.}
    \label{fig:predictions}
\end{figure*}

To further validate our method, we calculated cross sections for five reactions not included in the original optimization (Fig.~\ref{fig:predictions}). TALYS-recommended OMPs were used for all calculations. For each reaction, the optimized NLD parameters were applied only to the relevant nuclei from our optimization network, while default TALYS NLD and pre-equilibrium density $g$ parameters were assigned to all other nuclei.

Specifically, for the $^{64}$Zn$(\alpha,p)^{67}$Ga reaction, only the residual nucleus was part of our optimization. For the $^{66}$Zn$(p,n)^{66}$Ga and $^{66}$Zn$(\alpha,n)^{69}$Ge reactions, the predictions incorporate parameter variations for the target, compound, and residual nuclei, and for the latter reaction, the residual nucleus of the competing ($\alpha$,p) reaction was also part of the optimization. Finally, for the $^{65}$Cu$(p,\gamma)^{66}$Zn and $^{69}$Ga$(n,2n)^{68}$Ga reactions, both the target and residual nuclei were included in our optimization network.

The uncertainty envelopes (opaque teal bands) shown in Fig.~\ref{fig:predictions} were obtained from statistical model calculations using all Pareto-optimal parameter sets. For each reaction, the spread of cross-section values across all solutions defines the predicted uncertainty, while the most probable prediction (teal line) corresponds to the parameter set at the peak of the KDE distribution.

Our approach provides uncertainty-quantified cross sections that are mostly in fair agreement with existing data \cite{Levkovskij1991_a3n,Gyurky2012,Porile1959,Stelson_an,NAGAME1989,SZELECSENYI2005_a3n_u,Hermanne1999_a3n_u,Tarkanyi1990,Sevior1983,Raut2011,BORMANN1965,Luo2012,PU2003}. The uncertainty band reflects the propagated uncertainty from the optimized NLD parameters to the cross section of the “unknown” reactions. Further tuning of the optical potential could significantly improve the agreement at low energies. With these results, we show that our method can provide realistic cross-section estimates and address the community-identified need for consistent uncertainty quantification between model parameters and calculated observables.

\section{Model predictions}
\begin{figure}[thb]
    \includegraphics[width=.98\columnwidth]{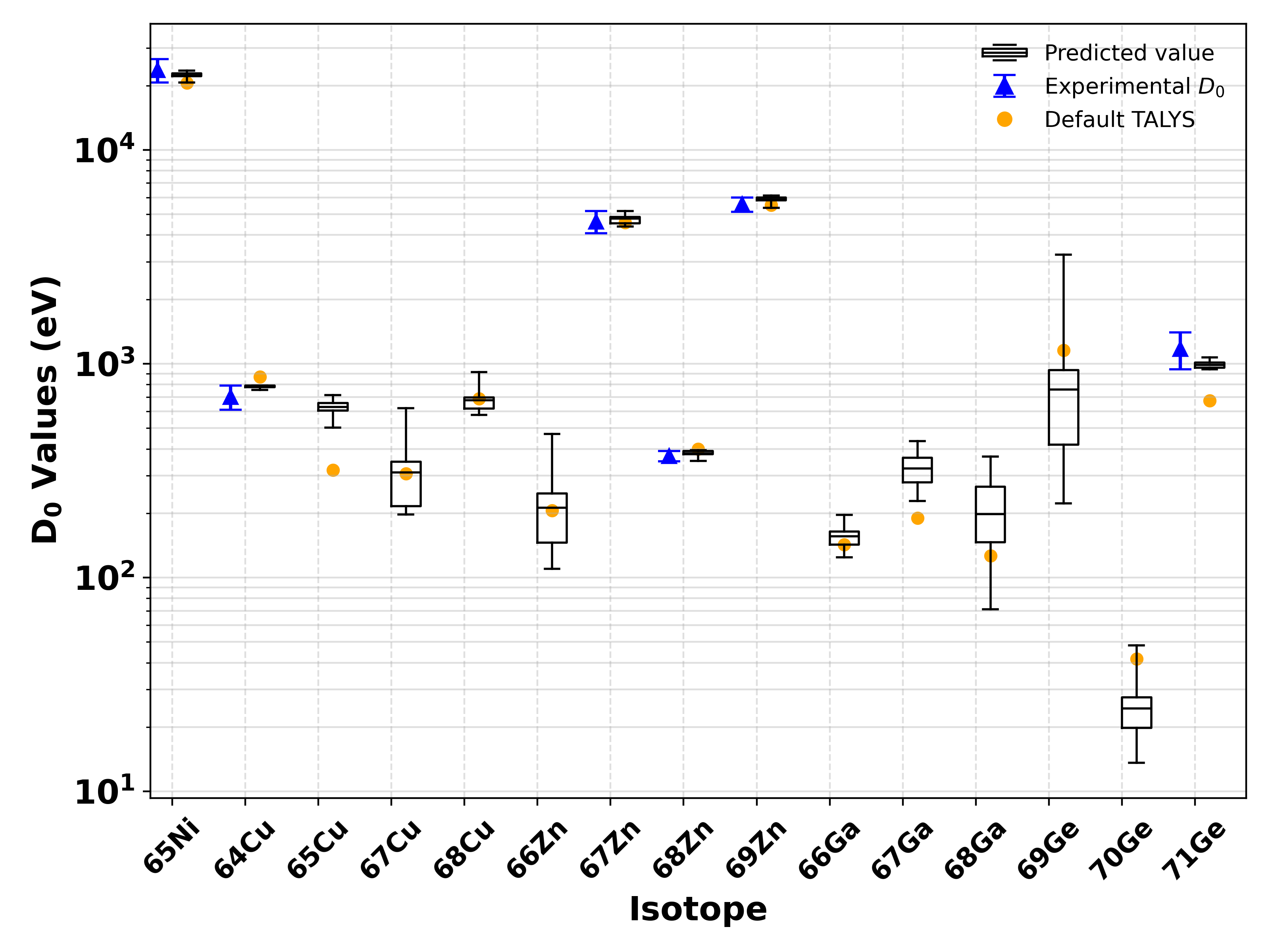} 
     \caption{Predicted $D_0$ values for each isotope in our calculation. Experimental values are shown as blue points, default TALYS (see text) predictions as orange points, and results from the optimization as black box plots. Boxes indicate the 68\% credible interval (16$^{th}$ - 84$^{th}$ percentile) representing the probable range, central lines indicate the most probable value (KDE peak), and whiskers denote the full range of values found in this work.}
    \label{fig:D0}
\end{figure}

Fig.~\ref{fig:D0} shows our uncertainty-quantified s-wave neutron spacings based on the optimized $\alpha(S_n)$ values of table \ref{tab:RESULTS} (open boxplot symbols).  
For comparison, we plot with blue color the available experimental data from the RIPL-3 parameter library that we used to constraint the calculations. With orange we represent any available RIPL-2 data (these are used by default in TALYS), or, when no data are available, the typical TALYS approach of a Fermi-gas level density model estimate using systematics and the available discrete levels.
Our calculations, in almost all cases, improve upon evaluated neutron resonance spacing uncertainties due to the additional constraints imposed by our optimization network. Furthermore, for the first time, we are able to provide uncertainty-quantified values for experimentally inaccessible cases. It has to be noted that we provide neutron resonance spacings for both sides of the valley of stability reaching two or three neutrons far from the last stable isotope of an element. The spacings obtained follow the expected systematic behavior, tracking the odd-even trends of the neutron separation energies. Moreover, and this is most important for those inaccessible cases, we quantify for the first time the effect of theoretical modeling quality to the estimated quantities and show how this uncertainty propagates differently to various isotopes. 
Our optimized level density parameters and associated data-driven uncertainties combined with existing low-lying discrete level information, effectively define the average energy dependence of the level density for all these isotopes. The included uncertainty, allows the use of this level density as an ingredient in uncertainty-quantified calculations of  observables like cross-sections and reaction rates that many physics applications need.

\section{Summary and Conclusions}
We have developed a data-driven optimization framework to extract uncertainty-quantified local parameterizations for statistical model inputs. By utilizing cross-section data from multiple reaction channels, the approach ensures consistency across all channels and incorporates competing experimental constraints into a coherent optimization scheme.
The optimized parameter sets accurately reproduce experimental data across all considered reaction channels, and exhibit improved agreement compared to results obtained with global TALYS 2.0 parameters and evaluated libraries. 

The optimized NLD parameters enabled the extraction of uncertainty-quantified $s$-wave resonance spacing $D_0$ values for nuclei a few neutrons either way of the stability valley. Experimental $D_0$ data are scarce even for stable isotopes, and entirely absent for neutron-deficient nuclei due to current experimental limitations. Our method thus constitutes the first reliable technique for estimating these important quantities with quantified uncertainties. Our results effectively define the average energy dependent level density for each of the isotopes studied with experimentally driven uncertainties.
Expanding our method to other areas of the nuclear chart has the potential to provide uncertainty-quantified statistical properties and facilitate reliable estimates of cross-sections for key nuclear physics applications.

\section{Acknowledgments}
This material is based upon work supported by the U.S. Department of Energy, Office of Science, Nuclear Physics program under Award Numbers DE-SC-0022538, DE-NA0004073, DE-FG02-88ER40387, and Central Michigan University College of Science and Engineering. It benefits from the LANSCE accelerator facility and is supported by the U.S. Department of Energy under contract No. 89233218CNA000001 and by the US Nuclear Data Program (USNDP) under the Office of Science of the U.S. Department of Energy.
% \newpage
% \clearpage
% \bibliography{BIB}

\end{document}